\documentclass[sigplan,screen]{acmart}
\AtBeginDocument{%
  }

\setcopyright{none}
\copyrightyear{2018}
\acmYear{2018}
\acmDOI{XXXXXXX.XXXXXXX}
\acmConference[Conference acronym 'XX]{Make sure to enter the correct
  conference title from your rights confirmation email}{June 03--05,
  2018}{Woodstock, NY}
\acmISBN{978-1-4503-XXXX-X/2018/06}

\usepackage{zi4}
\usepackage[shortlabels]{enumitem}
\usepackage{mathtools}
\usepackage{subcaption}
\usepackage{mathpartir}
\usepackage[frozencache]{minted}
\newminted{python}{
frame=lines,
framesep=2mm,
baselinestretch=0.9,
fontsize=\footnotesize,
tabsize=2,
numbersep=4pt,
vspace=0.5\topsep
}
\usepackage{cleveref}
\crefname{theorem}{Thm.}{Thms.}
\crefname{lemma}{Lem.}{Lemmas}
\crefname{corollary}{Cor.}{Cors.}
\crefname{figure}{Fig.}{Figs.}
\crefname{definition}{Defn.}{Defns.}
\crefname{table}{Tab.}{Tabs.}
\crefformat{section}{\S#2#1#3}
\crefmultiformat{section}{\S#2#1#3}{ and~\S#2#1#3}{, \S#2#1#3}{ and~\S#2#1#3}
\crefname{example}{Ex.}{Exs.}
\crefname{item}{item}{items}
\crefname{footnote}{footnote}{footnotes}
\crefname{observation}{Obs.}{Obs.}
\crefname{remark}{Remark}{Remarks}
\crefname{proposition}{Prop.}{Props.}
\crefname{equation}{Eqn.}{Eqns.}
\crefname{counterexample}{Counterexample}{Counterexamples}
\crefname{algorithm}{Algorithm}{Algorithms}
\crefname{principle}{Principle}{Principles}
\crefname{challenge}{Challenge}{Challenges}
\crefname{fact}{Fact}{Facts}
\usepackage{shortcuts}

\newcommand{\m}[1]{\mathsf{#1}}

\newcommand{\handler}[1]{\mathbf{handler}\; \{ #1 \}}
\newcommand{\mret}[1]{\mathbf{return}\; #1}
\newcommand{\mdo}[3]{\mathbf{let}\; #1 \leftarrow #2\; \mathbf{in}\; #3}
\newcommand{\mwith}[2]{\mathbf{with}\; #1\; \mathbf{do}\; #2}
\newcommand{\mif}[3]{\mathbf{if}\; #1\; \mathbf{then}\; #2\; \mathbf{else}\; #3}

\makeatletter
\newtheoremstyle{acmremark}%
  {.5\baselineskip\@plus.2\baselineskip\@minus.2\baselineskip}% space above
  {.5\baselineskip\@plus.2\baselineskip\@minus.2\baselineskip}% space below
  {\itshape}% body font
  {\z@}% indent amount
  {\itshape}% head font
  {.}% punctuation after head
  {.5em}% spacing after head
  {\thmname{#1}\thmnumber{ #2}\thmnote{ {\normalfont(#3)}}}% head spec
\makeatother
\AtEndPreamble{%
  \theoremstyle{acmplain}%
  \theoremstyle{acmdefinition}%
  \theoremstyle{acmremark}%
  \theoremstyle{acmplain}%
}
\AtBeginEnvironment{example}{%
  \pushQED{\qed}%
}
\AtEndEnvironment{example}{\popQED\endexample}

\AtEndPreamble{%
  \setlist{nosep,leftmargin=\parindent}%
}
\AtBeginDocument{%
  \setlength\abovedisplayskip{0.5\abovedisplayskip}%
  \setlength\belowdisplayskip{0.5\belowdisplayskip}%
  \setlength\abovedisplayshortskip{0.5\abovedisplayshortskip}%
  \setlength\belowdisplayshortskip{0.5\belowdisplayshortskip}%
  \setlength\floatsep{0.5\floatsep}%
  \setlength\abovecaptionskip{0.5\abovecaptionskip}%
}

\begin{document}

\title{Composable Effect Handling for Programming LLM-integrated Scripts}

\author{Di Wang}
\affiliation{
  \institution{Peking University}
  \country{China}
}


\begin{abstract}
% State the problem
% Say why it's an interesting problem
% Say what your solution achieves
% Say what follows from your solution

Implementing LLM-integrated scripts introduces challenges in modularity and performance, as scripts are often coupled to specific LLM implementations and fail to exploit parallelization opportunities.
This paper proposes using composable effect handling to separate workflow logic from effectful operations, such as LLM calls, I/O, and concurrency, enabling modularity without sacrificing the opportunity for performance optimization.
By treating these operations as abstract interfaces and discharging them via effect handlers, this paper shows that scripts can achieve significant speedups (e.g., 10$\times$ in a Tree-of-Thoughts case study) without compromising modularity.
This paper aims to promote composable effect handling as a programming style for LLM scripting. 
\end{abstract}





\maketitle

\section{Introduction}
\label{sec:intro}

% Three useful guides to the organization of the intro section
%
% Organization A (Simon Peyton-Jones)
%  o Describe the problem
%  o State your contributions
%  o Organization of the paper [if not able to cover point by point
%    in contributions]
%
% Organization B (Nickolai Zeldovich)
%  o Elevator story -- high-level statement of the problem
%  o The problem in more technical terms
%  o Conventional wisdom: sketch of previous techniques and their
%    shortcomings
%  o Describe the solution to the problem as a black box,
%    so that it is clear what our solution offers over previous work
%  o Technical challenges to obtaining a solution (e.g., 1 paragraph
%    for each)
%  o State how we solve the challenges (at most a few paragraphs)
%  o Experimental highlights
%  o Contributions
%  o Organization of the paper
%
% Organization C (Derek Dreyer)
%  o Context: Set the stage, motivate the general topic
%  o Gap: Explain your specific problem and why existing work does not
%    adequately solve it
%  o Innovation: State what you've done that is new, and explain how it
%    helps fill the gap

Over the past few years, an increasing number of developers have been involved in building intelligent workflows and agents powered by large language models (LLMs).
For example, consider the following Python script, which implements a simple workflow that asks an LLM for topics in a research area and then asks for a description for each topic:
\begin{pythoncode}
def research_topics():
	topics = get_topics("PL techniques for LLM applications")
	for topic in topics:
		log(topic)
		description = get_description(topic)
		log(description)
\end{pythoncode}
Here, both \verb|get_topics| and \verb|get_description| are external calls implemented as functions that wrap API calls to (possibly remote) LLM services.
Despite its simplicity, the script suffers from two non-trivial drawbacks:
\begin{itemize}
  \item \emph{Modularity:}
  The script is coupled with a concrete implementation of \verb|get_topics| and \verb|get_description|.
  This would break the modularity of programming when the developer wants to switch among different implementations of the LLM calls, or trace and mock them for testing.
  \item \emph{Performance:}
  The script takes a long time to execute: in a sample execution, each LLM call took approximately 2 seconds to run, and the entire script took around 20 seconds.
  The performance is not ideal because the script does not exploit the LLM service's capacity for parallelization.
\end{itemize}
One way to maintain modularity is by using object-oriented programming.
For example, the developer can declare an interface with methods \verb|get_topics| and \verb|get_description| and parameterize the script by a concrete instantiation of the interface.
Using inheritance and polymorphism, the developer can modularly create instantiations for executing, tracking, or mocking LLM calls.

However, it would be more challenging to improve performance while maintaining modularity.
For example, in the script above, the description of each topic could be requested in parallel, and exploiting such parallelism could result in a total runtime of about 5 seconds, achieving a 4$\times$ speedup.
Unfortunately, it typically requires developers to modify the code substantially and carefully to utilize parallelism.
Moreover, it would ``pollute'' the script with explicit parallel or asynchronous primitives (e.g., Python's \mintinline{python}{async} and \mintinline{python}{await}) and thus break modularity.

In this paper, we propose to use \emph{composable effect handling}~\cite{FOSSACS:PP01,ESOP:PP09} for the modular programming of efficient LLM-integrated scripts, like the one shown at the beginning.
With our approach, simply wrapping the workflow script with a few effect handlers yields an efficient implementation:
\begin{pythoncode}
with (
	# handler for async operations
	AsyncHandler(),
	# handler for LLM-call operations
	AsyncLLMHandler(**llm_kwargs),
	# handler for rendering async side effects sequentially
	AsyncSeqHandler(),
	# handler for application-specific operations
	AsyncResearchTopicsHandler(),
):
	research_topics() # the workflow script stays the same!
\end{pythoncode}
%
%A synchronous implementation:
%%
%\begin{pythoncode}
%with (
%	# handler for LLM-call operations
%	LLMHandler(**llm_kwargs),
%	# handler for application-specific operations
%	ResearchTopicsHandler(),
%):
%	research_topics() # the workflow script stays the same!
%\end{pythoncode}
%
Or an implementation that mocks LLM calls for testing:
\begin{pythoncode}
with (
	AsyncHandler(),
	# handler for mocking LLM-call operations using a trace
	AsyncReplayLLMHandler(trace),
	AsyncSeqHandler(), AsyncResearchTopicsHandler(),
):
	research_topics() # the workflow script stays the same!
\end{pythoncode}

In this paper, we make the following contributions:
\begin{itemize}
  \item
  We describe a paradigm that uses abstract operations and effect handlers for LLM scripting (\cref{sec:approach}).
  Our approach achieves the separation of workflow logic and effects, including LLM calls, I/O, and concurrency.
%  and even nondeterminism.
  %
%  Notably, our approach does not rely on (and could be integrated with) any library or framework for LLM programming (e.g., \cite{misc:LangChain25,misc:Agno25,ICLR:KSM24,arxiv:ZYX24,misc:Guidance25,misc:LlamaIndex25,misc:OpenAIAgents25,PLDI:BFV23,AAAI:LHL24}).
  %
  \item
  We present a case study on implementing Tree-of-Thoughts (ToT)~\cite{NIPS:YYZ23} (\cref{sec:case-study}).
  %
%  ToT treats problem solving and reasoning as a graph search problem, featuring a more complex workflow logic than the simple script shown at the beginning.
  %
%  ToT's official implementation~\cite{misc:ToT25} does not use parallelization.
  %
  With our approach, one can implement ToT's workflow in a sequential and clean manner, while being able to implement handlers separately to exploit parallelism and achieve a 10$\times$ speedup.
\end{itemize}
\section{Approach}
\label{sec:approach}

\paragraph{Background}
Following the terminology of \citet{ICFP:KLO13}, we view composable effect handling as two parts:
\begin{itemize}
  \item \emph{Abstract operations:}
  \citet{FOSSACS:PP01} introduced the algebraic theory of computational effects.
  This suggests that we can treat effectful functions---including \verb|get_topics|, \verb|get_description|, and even \verb|log|---as abstract operations.
  These abstract operations serve as interfaces that separate effects from computations.
  Abstract operations can compose, enabling \emph{modular abstraction}.
\begin{pythoncode}
# an operation is a callable object
get_topics = Operation()
get_description = Operation()
log = Operation()
\end{pythoncode}
  \item \emph{Effect handlers:}
  To determine the behavior of abstract operations, \citet{ESOP:PP09} introduced effect handlers.
  An effect handler discharges a particular set of abstract operations, leaving others still abstract to be further discharged by other handlers.
  Effect handlers can also compose, enabling \emph{modular instantiation}.
\begin{pythoncode}
# a handler is a manager for a set of operations
class LogHandler(Handler):
	def __init__(self):
		super().__init__()
		# this handler discharges the `log` operation
		self.register(log, self.log)

	def log(self, msg): print(f"[INFO] {msg}")
\end{pythoncode}
\end{itemize}

Operationally, the runtime should maintain a stack of active effect handlers (e.g., through Python's \mintinline{python}{with} blocks) and dispatch an abstract operation to the topmost handler that can discharge this operation in the stack.
Note that executing the operation itself may further invoke other operations, which are supposed to be discharged in a sub-stack below the current handler.
For example, below implements another handler for the \verb|log| operation to also print the current time:
\begin{pythoncode}
class LogDateHandler(Handler):
	def __init__(self):
		super().__init__()
		self.register(log, self.log)

	def log(self, msg):
		print(f"[DATE] {datetime.now()}")
		# this handler invokes the `log` operation
		log(msg)  
\end{pythoncode}
In this way, \verb|LogDateHandler| can be composed with any handlers that discharge \verb|log|.
Composing \verb|LogHandler| and \verb|LogDateHandler| refines any computation that invokes \verb|log| to also print the current time:
\begin{pythoncode}
with LogHandler(), LogDateHandler(): log("Hello World!")
# [DATE] 2025-06-29 20:25:17.102486 
# [INFO] Hello World!
\end{pythoncode}

Note that all the handlers considered so far are \emph{one-shot}~\cite{TFP:KK20}, i.e., the handler resumes the computation from invoking an operation \emph{once}.
Effect handlers can be \emph{multi-shot}~\cite{ENTCS:Pretnar15,ICFP:KLO13,JLAMP:BP15}, enabling more complex forms of effects such as nondeterminism.
We will focus on one-shot handlers in this paper and discuss multi-shot ones in the Appendix (\cref{sec:multi-shot}).

\paragraph{Abstract Operations}
In this paper, we consider four abstract operations to demonstrate LLM scripting:
\begin{itemize}
  \item \verb|async_(coro, post_fn)|
  is intended to schedule a coroutine \verb|coro| and return a future object that represents the eventual result of \verb|coro|.
  The operation takes an optional argument \verb|post_fn| as a callback function on the result.
  \item \verb|await_(fut)|
  is intended to wait until a future object \verb|fut| becomes available and return its result.
  \item \verb|complete(prompt)|
  is intended to request an LLM service to generate text from a \verb|prompt| and return the output text.
  \item \verb|parse(prompt, schema)|
  is intended to request an LLM service to generate a structured output from a \verb|prompt| and a \verb|schema|, and return the output as a parsed object.
\end{itemize}
Note that these operations are not coupled with any concrete implementation:
\verb|async_| and \verb|await_| can be discharged by different concurrent mechanisms;
\verb|complete| and \verb|parse| can be synchronous or asynchronous, depending on their handler;
and in the asynchronous case, \verb|complete| and \verb|parse| are supposed to return a future object.

We now describe how to compose these operations to implement the \verb|get_topics|, \verb|get_description|, and \verb|log| for the LLM script \verb|research_topics| shown at the beginning of \cref{sec:intro}.
We start with the \verb|get_topics| operation, which requests the LLM service to return a list of research topics:
\begin{pythoncode}
def get_topics(area):
	return await_(parse(
		f"Give a list of topics in the research area {area}.",
		ResearchArea, # a schema with a field `topics`
	)).topics  
\end{pythoncode}
We then consider the \verb|get_description| operation, which requests LLM to describe a topic.
To exploit parallelism, we do not wait for the completion of this operation so that it would return a future object that wraps the description:
\begin{pythoncode}
def get_description(topic):
	return complete(
		f"Give a short description about the topic {topic}."
	)
\end{pythoncode}
Finally, we need to implement the \verb|log| operation to account for the case where its argument is a future object: in line 6 of the script shown at the beginning of \cref{sec:intro}, \verb|log| is invoked with a future object resulted from \verb|get_description|:
\begin{pythoncode}
def log(msg):
	async def aux(): # use `async` to define a coroutine
		return await msg if isinstance(msg, Awaitable) else msg
	return async_(aux(), print) # use `print` as a callback
\end{pythoncode}
%
%In our implementation, these operations are collected in an effect handler \verb|AsyncResearchTopicsHandler|.

\paragraph{Effect Handlers}
We now sketch the implementation of \verb|AsyncHandler|, \verb|AsyncLLMHandler|, and \verb|AsyncSeqHandler|.

For \verb|AsyncHandler|, we use Python's \verb|asyncio| library to discharge the \verb|async_| and \verb|await_| operations.
When entering the handler (i.e., when the handler is placed on the effect-handler stack), it creates an event loop to schedule coroutines.
When exiting the handler (i.e., when the handler is popped from the stack), it runs all scheduled coroutines to completion and then closes the loop.
Below is the code:
\begin{pythoncode}
class AsyncHandler(Handler): # handles async_, await_
	def __enter__(self):
		super().__enter__()
		self.loop = asyncio.new_event_loop()
		self.futures = []
		return self

	def __exit__(self, *exc):
		for future in self.futures:
			self.loop.run_until_complete(future)
		self.loop.close()
		return super().__exit__(*exc)
\end{pythoncode}
The \verb|async_(coro, post_fn)| operation creates a task for the coroutine \verb|coro|.
It does not wait for the completion of the coroutine and returns a future object immediately:
\begin{pythoncode}
	def async_(self, coro, post_fn=lambda x: x):
		async def _coro():
			result = await coro
			return post_fn(result)
		future = self.loop.create_task(_coro())
		self.futures.append(future)
		return future
\end{pythoncode}
The \verb|await_(future)| operation waits for the completion of the future object \verb|future| and then returns the result:
\begin{pythoncode}
	def await_(self, future):
		self.loop.run_until_complete(future)
		return future.result()
\end{pythoncode}

For \verb|AsyncLLMHandler|, we use OpenAI's official Python library~\cite{misc:OpenAIPython25}, but one can use any library for requesting LLM.
The handler discharges \verb|complete| and \verb|parse|, and it further invokes the \verb|async_|.
Because OpenAI's library supports asynchronous usage, the implementation is straightforward:
\begin{pythoncode}
class AsyncLLMHandler(Handler): # handles complete, parse
	def complete(self, prompt):
		async def aux():
			r = await self.client.chat.completions.create(...)
			return r.choices[0].message.content
		return async_(aux())

	def parse(self, prompt, schema):
		async def aux():
			r = await self.client.beta.chat.completions.parse(...)
			return r.choices[0].message.parsed
		return async_(aux())
\end{pythoncode}

Finally, \verb|AsyncSeqHandler| is an interesting handler because it is intended to refine the behavior of \verb|async_|:
the callbacks (\verb|post_fn|'s) of coroutines should be executed sequentially.
This is useful if the callbacks generate observable side effects, e.g., printing information as in \mintinline{python}|async_(aux(), print)|.
%
%In our implementation,
This handler both discharges and invokes \verb|async_|.
It creates a new event for each coroutine to signal the completion of the coroutine, and enforces its callback to wait for the previous coroutine's signal event:
\begin{pythoncode}
class AsyncSeqHandler(Handler): # handles async_
	def __enter__(self):
		super().__enter__()
		self.init_event = asyncio.Event()
		self.prev_event = self.init_event
		self.init_event.set()
		return self

	def async_(self, coro, post_fn=lambda x: x):
		async def _coro(prev_event, next_event):
			result = await coro
			await prev_event.wait()  # wait for the previous event
			result = post_fn(result)
			next_event.set()         # trigger the next event
			return result
		next_event = asyncio.Event()
		future = async_(_coro(self.prev_event, next_event))
		self.prev_event = next_event
		return future  
\end{pythoncode}

By bringing all the ingredients together, we enable composable effect handling for the modular programming of efficient LLM-integrated scripts, as highlighted in \cref{sec:intro}.
%
%\begin{pythoncode}
%with (
%	AsyncHandler(), AsyncLLMHandler(**llm_kwargs),
%	AsyncSeqHandler(), AsyncResearchTopicsHandler(),
%):
%	topics = get_topics("PL techniques for LLM applications")
%	for topic in topics:
%		log(topic)
%		description = get_description(topic)
%		log(description)
%\end{pythoncode}
%
%Moreover, general-purpose handlers such as \verb|AsyncHandler| and \verb|AsyncLLMHandler| can be reused for other LLM scripting tasks.
\section{Case Study: Tree-of-Thoughts}
\label{sec:case-study}

We consider Tree-of-Thoughts (ToT)~\cite{NIPS:YYZ23} as a case study to evaluate our composable-effect-handling approach.
ToT generalizes Chain-of-Thought (CoT) by transforming logical reasoning from following a \emph{chain} of thoughts into exploring a \emph{tree} of thoughts, where a thought represents an intermediate step for solving a problem.
ToT treats problem solving as searching through a graph, where the nodes and edges represent partial-solve states and thoughts, respectively.
The ``tree'' is then a search tree through the graph, e.g., a depth-first or breadth-first search tree.
ToT relies on LLMs to
(i) generate thoughts, i.e., expand a partial-solve state to get multiple candidates for the next state, and
(ii) evaluate these candidates, i.e., assign a fitness score to each candidate.

Below is our implementation of ToT with beam search, where \verb|init|, \verb|expand|, and \verb|score| are  operations for getting the inital solve state, generating thoughts, and evaluating candidates, respectively; \verb|n_steps|, \verb|n_select|, and \verb|n_eval| are the number of search steps, states kept in the frontier, and repetitions for evaluating a candidate, respectively:
\begin{pythoncode}
init, expand, score = Operation(), Operation(), Operation()

def tree_of_thoughts(n_steps, n_select, n_eval):
	frontier = [init()]
	for _ in range(n_steps):
		expanded = [expand(state) for state in frontier]
		candidates = chain(*expanded)         # flatten the list
		scored = [score(cand, n_eval) for cand in candidates]
		frontier = top_k(scored, n_select)    # select greedily
	print(frontier)
\end{pythoncode}
In each step, the script invokes \verb|expand| to generate a list of candidates for each state in the frontier, then invokes \verb|score| to calculate a score for each candidate, and finally keeps at most \verb|n_select| candidates with the highest scores in the frontier.
Note that \verb|chain| and \verb|top_k| are normal functions.

We then implement \verb|AsyncGame24Handler|, an effect handler that discharges \verb|expand| and \verb|score| for Game of 24, whose reasoning goal is to use four numbers and elementary operators to obtain 24.
Following the setup of ToT's paper~\cite{NIPS:YYZ23}, we use a sequence of equations (e.g., ``[10 - 2 = 8]'' and ``[10 - 2 = 8; 13 - 10 = 3]'') and a list of remaining numbers (e.g., ``[8; 10; 13]'') to represent a partial-solve state.
We implement \verb|expand| by asking the LLM to generate possible next equations from remaining numbers (e.g., ``8 + 10 = 18'' and ``13 - 10 = 3'').
We implement \verb|score| by asking the LLM to determine whether remaining numbers (e.g., ``[18; 13]'' and ``[8; 3]'') are ``sure/likely/impossible'' to obtain 24.
Composing the general \verb|tree_of_thoughts| script and the handler for Game of 24 yields an efficient script for solving Game of 24 via ToT:
\begin{pythoncode}
with (
	AsyncHandler(), AsyncLLMHandler(**llm_kwargs),
	AsyncGame24Handler(),
):
	tree_of_thoughts(n_steps=4, n_select=5, n_eval=3)
\end{pythoncode}
Note that we reuse \verb|AsyncHandler| and \verb|AsyncLLMHandler| presented in \cref{sec:approach}.
We set the number of steps to 4 because we add a final step to ask the LLM to validate the solution and extract a single expression (e.g., ``(10 - 2) * (13 - 10) = 24'').

We compared the running time of our implementation on three inputs against a synchronous version that is conceptually similar to the ToT's official implementation~\cite{misc:ToT25}.
We conducted our evaluation on a machine equipped with an Apple M2 processor (3.50 GHz, eight cores) and 24 GB of RAM, running macOS Sequoia 15.5.
For remote LLM calls, we used Qwen-Turbo~\cite{arxiv:Qwen25}.
\cref{tab:eval} presents the statistics, indicating that our approach achieves an average speedup of 10.88$\times$.
Notably, to obtain the synchronous version, we simply replace \verb|AsyncGame24Handler| with another effect handler \verb|Game24Handler|, demonstrating good modularity.

\begin{table}[t]
\caption{Running time (seconds) on several inputs.}
\label{tab:eval}
\begin{tabular}{l|r|r|r}
  \hline
  \textbf{Input} & \textbf{Async} & \textbf{Sync} & \textbf{Speedup} \\ \hline
  {[4; 9; 10; 13]} & 36.93 & 412.57 & 11.17$\times$ \\
  {[2; 10; 10; 13]} & 34.16 & 458.03 & 13.41$\times$ \\
  {[5; 6; 8; 13]} & 35.01 & 464.54 & 13.27$\times$ \\ \hline
\end{tabular}
\vspace{-0.5em}
\end{table}
\section{Related Work}
\label{sec:related}

%\paragraph{Libraries and Frameworks for LLM Programming}
%
Recent work has proposed many libraries and frameworks to aid programming LLM-integrated scripts.
We take inspiration from \textsc{Epic}~\cite{arxiv:MKZ25}, a Python library that provides an opportunistic evaluation strategy for LLM scripts with standard control-flow constructs.
\textsc{Epic} supports a runtime environment for this evaluation strategy, which automatically parallelizes independent LLM calls (as well as other external calls) and streams the results of the calls.
With such runtime support, \textsc{Epic} achieves significant performance gains by exploiting parallelism without disrupting the workflow logic.
However, it is unclear whether \textsc{Epic} enjoys reasonable modularity, as the workflow logic is coupled with its runtime.
Our work takes an orthogonal approach, treating external calls as abstract operations to achieve modular abstraction.
We use effect handlers to discharge these operations, thereby achieving modular instantiation.
Implementing \textsc{Epic}'s evaluation strategy as an effect handler could be done in our approach; we leave this for future work.

Most existing libraries and frameworks aim to enhance the modularity of LLM scripting,
including
LangChain~\cite{misc:LangChain25},
Agno~\cite{misc:Agno25},
DSPy~\cite{ICLR:KSM24},
SGLang~\cite{arxiv:ZYX24},
Guidance~\cite{misc:Guidance25},
LlamaIndex~\cite{misc:LlamaIndex25},
OpenAI Agents~\cite{misc:OpenAIAgents25},
LMQL~\cite{PLDI:BFV23},
Vieira~\cite{AAAI:LHL24},
etc.
They serve as \emph{domain-specific languages} (DSLs) that provide modular interfaces for specific aspects of LLM scripting, such as retrieval-augmented generation, advanced prompt engineering, multi-agent workflows, and interoperability.
Our work, on the other hand, aims to promote a \emph{programming style}, namely composable effect handling, for LLM scripting.
Integrating our approach with these DSLs is theoretically feasible; we leave this for future work.

%\paragraph{Effect Handlers for Probabilistic Programming}
%
We are not aware of any prior work on adapting effect handlers to LLM scripting.
Nevertheless, we take inspiration from recent adaptations in the field of probabilistic programming.
Probabilistic programming languages (PPLs) aim to provide an interface between complex probabilistic models and advanced inference algorithms.
\citet{HOPE:SK15} proposed and \citet{HASKELL:NPW23} advanced the idea of structuring inference algorithms as \emph{inference patterns} (i.e., computations in terms of inference-related abstract operations) and \emph{pattern instantiations} (i.e., effect handlers for these abstract operations).
Two popular PPLs---Pyro~\cite{JMLR:BCJ18} and Edward~\cite{ICLR:THS17,arxiv:MG18}---benefit from and justify the practicality of the effect-handler-based design.
Given recent studies on combining probabilistic programming with LLMs~\cite{arxiv:LZG24,ICLR:HJE24,ICML:ZBM24}, it is interesting future work to unify our effect handlers for LLM scripting and those for probabilistic programming.
\section{Conclusion}
\label{sec:concl}

\newenvironment{myquote}%
  {\list{}{\leftmargin=0.25in\rightmargin=0.15in}\item[]}%
  {\endlist}

In this paper, we present a programming style that facilitates composable effect handling for programming LLM-integrated scripts.
Structuring LLM scripts---including external calls like LLM calls---via abstract operations and effect handlers enables modular workflow programming, without sacrificing the potential for performance optimization.
On a broader level, we adhere to the philosophy proposed in a blog post~\cite{misc:BuildingAgents24} by Anthropic, the company behind Claude:
\begin{myquote}
  ``The most successful implementations weren't using complex frameworks or specialized libraries. Instead, they were building with simple, composable patterns.''
\end{myquote}

%Our work remains preliminary in several aspects.
%%
%Firstly, we have not fully explored the capability of effect handlers, especially multi-shot ones, for programming LLM scripts that perform backtracking or more general searching.
%%
%Secondly, we have not conducted a thorough empirical evaluation of our approach on a diverse suite of LLM scripts, such as retrieval-augmented generation, constrained decoding, tool calling, and memory, among others.
%%
%Thirdly, we have not attempted to integrate our approach with existing libraries and frameworks for LLM programming to assess its versatility.
%%
%These aspects make up our future work.

%\begin{acks}
%To Robert, for the bagels and explaining CMYK and color spaces.
%\end{acks}

\bibliographystyle{ACM-Reference-Format}
\bibliography{db}


\begin{thebibliography}{29}

%%% the Web address.

\ifx \showCODEN    \undefined \def \showCODEN     #1{\unskip}     \fi
\ifx \showISBNx    \undefined \def \showISBNx     #1{\unskip}     \fi
\ifx \showISBNxiii \undefined \def \showISBNxiii  #1{\unskip}     \fi
\ifx \showISSN     \undefined \def \showISSN      #1{\unskip}     \fi
\ifx \showLCCN     \undefined \def \showLCCN      #1{\unskip}     \fi
\ifx \shownote     \undefined \def \shownote      #1{#1}          \fi
\ifx \showarticletitle \undefined \def \showarticletitle #1{#1}   \fi
\ifx \showURL      \undefined \def \showURL       {\relax}        \fi
% invisible to TeX
\providecommand\bibfield[2]{#2}
\providecommand\bibinfo[2]{#2}
\providecommand\natexlab[1]{#1}
\providecommand\showeprint[2][]{arXiv:#2}

\bibitem[{Agno, Inc.}(2025)]%
        {misc:Agno25}
\bibfield{author}{\bibinfo{person}{{Agno, Inc.}}} \bibinfo{year}{2025}\natexlab{}.
\newblock \bibinfo{title}{{agno-ai/agno: Full-stack framework for building Multi-Agent Systems with memory, knowledge and reasoning.}}
\newblock \bibinfo{howpublished}{Available on \url{https://github.com/agno-agi/agno}}.
\newblock


\bibitem[{Anthropic}(2024)]%
        {misc:BuildingAgents24}
\bibfield{author}{\bibinfo{person}{{Anthropic}}.} \bibinfo{year}{2024}\natexlab{}.
\newblock \bibinfo{title}{{Building effective agents}}.
\newblock \bibinfo{howpublished}{Available on \url{https://www.anthropic.com/engineering/building-effective-agents}}.
\newblock


\bibitem[Bauer and Pretnar(2015)]%
        {JLAMP:BP15}
\bibfield{author}{\bibinfo{person}{Andrej Bauer} {and} \bibinfo{person}{Matija Pretnar}.} \bibinfo{year}{2015}\natexlab{}.
\newblock \showarticletitle{{Programming with algebraic effects and handlers}}.
\newblock \bibinfo{journal}{\emph{J.\ Logical and Algebraic Methods in Programming}}  \bibinfo{volume}{84} (\bibinfo{date}{January} \bibinfo{year}{2015}), \bibinfo{pages}{108--123}.
\newblock
Issue 1.
\href{https://doi.org/10.1016/j.jlamp.2014.02.001}{doi:\nolinkurl{10.1016/j.jlamp.2014.02.001}}


\bibitem[Beurer-Kellner et~al\mbox{.}(2023)]%
        {PLDI:BFV23}
\bibfield{author}{\bibinfo{person}{Luca Beurer-Kellner}, \bibinfo{person}{Marc Fischer}, {and} \bibinfo{person}{Martin Vechev}.} \bibinfo{year}{2023}\natexlab{}.
\newblock \showarticletitle{{Prompting Is Programming: A Query Language for Large Language Models}}.
\newblock \bibinfo{journal}{\emph{Proc.\ ACM Program.\ Lang.}} \bibinfo{volume}{7}, \bibinfo{number}{186} (\bibinfo{date}{June} \bibinfo{year}{2023}), \bibinfo{pages}{1946--1969}.
\newblock
Issue PLDI.
\href{https://doi.org/10.1145/3591300}{doi:\nolinkurl{10.1145/3591300}}


\bibitem[Bingham et~al\mbox{.}(2018)]%
        {JMLR:BCJ18}
\bibfield{author}{\bibinfo{person}{Eli Bingham}, \bibinfo{person}{Jonathan~P. Chen}, \bibinfo{person}{Martin Jankowiak}, \bibinfo{person}{Fritz Obermeyer}, \bibinfo{person}{Neeraj Pradhan}, \bibinfo{person}{Theofanis Karaletsos}, \bibinfo{person}{Rishabh Singh}, \bibinfo{person}{Paul Szerlip}, \bibinfo{person}{Paul Horsfall}, {and} \bibinfo{person}{Noah~D. Goodman}.} \bibinfo{year}{2018}\natexlab{}.
\newblock \showarticletitle{{Pyro: Deep Universal Probabilistic Programming}}.
\newblock \bibinfo{journal}{\emph{J.\ Machine Learning Research}}  \bibinfo{volume}{20} (\bibinfo{date}{January} \bibinfo{year}{2018}).
\newblock
Issue 1.
\urldef\tempurl%
\url{https://dl.acm.org/doi/10.5555/3322706.3322734}
\showURL{%
\tempurl}


\bibitem[Hu et~al\mbox{.}(2024)]%
        {ICLR:HJE24}
\bibfield{author}{\bibinfo{person}{Edward~J. Hu}, \bibinfo{person}{Moksh Jain}, \bibinfo{person}{Eric Elmoznino}, \bibinfo{person}{Younesse Kaddar}, \bibinfo{person}{Guillaume Lajoie}, \bibinfo{person}{Yoshua Bengio}, {and} \bibinfo{person}{Nikolay Malkin}.} \bibinfo{year}{2024}\natexlab{}.
\newblock \showarticletitle{{Amortizing intractable inference in large language models}}. In \bibinfo{booktitle}{\emph{Int.\ Conf.\ on Learning Representations}} \emph{(\bibinfo{series}{ICLR'24})}.
\newblock


\bibitem[Kammar et~al\mbox{.}(2013)]%
        {ICFP:KLO13}
\bibfield{author}{\bibinfo{person}{Ohad Kammar}, \bibinfo{person}{Sam Lindley}, {and} \bibinfo{person}{Nicolas Oury}.} \bibinfo{year}{2013}\natexlab{}.
\newblock \showarticletitle{{Handlers in Action}}. In \bibinfo{booktitle}{\emph{Int.\ Conf.\ on Functional Programming}} \emph{(\bibinfo{series}{ICFP'13})}. \bibinfo{pages}{145--158}.
\newblock
\href{https://doi.org/10.1145/2500365.2500590}{doi:\nolinkurl{10.1145/2500365.2500590}}


\bibitem[Kawahara and Kameyama(2020)]%
        {TFP:KK20}
\bibfield{author}{\bibinfo{person}{Satoru Kawahara} {and} \bibinfo{person}{Yukiyoshi Kameyama}.} \bibinfo{year}{2020}\natexlab{}.
\newblock \showarticletitle{{One-shot Algebraic Effects as Coroutines}}. In \bibinfo{booktitle}{\emph{Trends in Functional Programming}} \emph{(\bibinfo{series}{TFP'20})}. \bibinfo{pages}{159--179}.
\newblock
\href{https://doi.org/10.1007/978-3-030-57761-2_8}{doi:\nolinkurl{10.1007/978-3-030-57761-2_8}}


\bibitem[Khattab et~al\mbox{.}(2024)]%
        {ICLR:KSM24}
\bibfield{author}{\bibinfo{person}{Omar Khattab}, \bibinfo{person}{Arnav Singhvi}, \bibinfo{person}{Paridhi Maheshwari}, \bibinfo{person}{Zhiyuan Zhang}, \bibinfo{person}{Keshav Santhanam}, \bibinfo{person}{Sri Vardhamanan}, \bibinfo{person}{Saiful Haq}, \bibinfo{person}{Ashutosh Sharma}, \bibinfo{person}{Thomas~T. Joshi}, \bibinfo{person}{Hanna Moazam}, \bibinfo{person}{Heather Miller}, \bibinfo{person}{Matei Zaharia}, {and} \bibinfo{person}{Christopher Potts}.} \bibinfo{year}{2024}\natexlab{}.
\newblock \showarticletitle{{DSPy: Compiling Declarative Language Model Calls into Self-Improving Pipelines}}. In \bibinfo{booktitle}{\emph{Int.\ Conf.\ on Learning Representations}} \emph{(\bibinfo{series}{ICLR'24})}.
\newblock


\bibitem[{LangChain, Inc.}(2025)]%
        {misc:LangChain25}
\bibfield{author}{\bibinfo{person}{{LangChain, Inc.}}} \bibinfo{year}{2025}\natexlab{}.
\newblock \bibinfo{title}{{langchain-ai/langchain: Build context-aware reasoning applications}}.
\newblock \bibinfo{howpublished}{Available on \url{https://github.com/langchain-ai/langchain}}.
\newblock


\bibitem[Lew et~al\mbox{.}(2024)]%
        {arxiv:LZG24}
\bibfield{author}{\bibinfo{person}{Alexander~K. Lew}, \bibinfo{person}{Tan Zhi-Xuan}, \bibinfo{person}{Gabriel Grand}, {and} \bibinfo{person}{Vikash~K. Mansinghka}.} \bibinfo{year}{2024}\natexlab{}.
\newblock \bibinfo{title}{{Sequential Monte Carlo Steering of Large Language Models using Probabilistic Programs}}.
\newblock
\href{https://doi.org/abs/2306.03081}{doi:\nolinkurl{abs/2306.03081}}


\bibitem[Li et~al\mbox{.}(2024)]%
        {AAAI:LHL24}
\bibfield{author}{\bibinfo{person}{Ziyang Li}, \bibinfo{person}{Jiani Huang}, \bibinfo{person}{Jason Liu}, \bibinfo{person}{Felix Zhu}, \bibinfo{person}{Eric Zhao}, \bibinfo{person}{William Dodds}, \bibinfo{person}{Neelay Velingker}, \bibinfo{person}{Rajeev Alur}, {and} \bibinfo{person}{Mayur Naik}.} \bibinfo{year}{2024}\natexlab{}.
\newblock \showarticletitle{{Relational Programming with Foundation Models}}. In \bibinfo{booktitle}{\emph{AAAI Conf.\ on Artif.\ Intelligence}} \emph{(\bibinfo{series}{AAAI'24})}. \bibinfo{pages}{10635--10644}.
\newblock
\href{https://doi.org/10.1609/aaai.v38i9.28934}{doi:\nolinkurl{10.1609/aaai.v38i9.28934}}


\bibitem[Liu(2025)]%
        {misc:LlamaIndex25}
\bibfield{author}{\bibinfo{person}{Jerry Liu}.} \bibinfo{year}{2025}\natexlab{}.
\newblock \bibinfo{title}{{run-llama/llama\_index: LlamaIndex is the leading framework for building LLM-powered agents over your data.}}
\newblock \bibinfo{howpublished}{Available on \url{https://github.com/run-llama/llama\_index}}.
\newblock


\bibitem[Mell et~al\mbox{.}(2025)]%
        {arxiv:MKZ25}
\bibfield{author}{\bibinfo{person}{Stephen Mell}, \bibinfo{person}{Konstantinos Kallas}, \bibinfo{person}{Steve Zdancewic}, {and} \bibinfo{person}{Osbert Bastani}.} \bibinfo{year}{2025}\natexlab{}.
\newblock \bibinfo{title}{{Opportunistically Parallel Lambda Calculus}}.
\newblock
\urldef\tempurl%
\url{https://arxiv.org/abs/2405.11361}
\showURL{%
\tempurl}


\bibitem[Moore and Gorinova(2018)]%
        {arxiv:MG18}
\bibfield{author}{\bibinfo{person}{Dave Moore} {and} \bibinfo{person}{Maria~I. Gorinova}.} \bibinfo{year}{2018}\natexlab{}.
\newblock \bibinfo{title}{{Effect Handling for Composable Program Transformations in Edward2}}.
\newblock
\urldef\tempurl%
\url{https://arxiv.org/abs/1811.06150}
\showURL{%
\tempurl}


\bibitem[Nguyen et~al\mbox{.}(2023)]%
        {HASKELL:NPW23}
\bibfield{author}{\bibinfo{person}{Minh Nguyen}, \bibinfo{person}{Roly Perera}, \bibinfo{person}{Meng Wang}, {and} \bibinfo{person}{Steven Ramsay}.} \bibinfo{year}{2023}\natexlab{}.
\newblock \showarticletitle{{Effect Handlers for Programmable Inference}}. In \bibinfo{booktitle}{\emph{Symp.\ on Haskell}} \emph{(\bibinfo{series}{Haskell'23})}. \bibinfo{pages}{44--58}.
\newblock
\href{https://doi.org/10.1145/3609026.3609729}{doi:\nolinkurl{10.1145/3609026.3609729}}


\bibitem[{OpenAI}(2025a)]%
        {misc:OpenAIAgents25}
\bibfield{author}{\bibinfo{person}{{OpenAI}}.} \bibinfo{year}{2025}\natexlab{a}.
\newblock \bibinfo{title}{{openai/openai-agents-python: A lightweight, powerful framework for multi-agent workflows}}.
\newblock \bibinfo{howpublished}{Available on \url{https://github.com/openai/openai-agents-python}}.
\newblock


\bibitem[{OpenAI}(2025b)]%
        {misc:OpenAIPython25}
\bibfield{author}{\bibinfo{person}{{OpenAI}}.} \bibinfo{year}{2025}\natexlab{b}.
\newblock \bibinfo{title}{{openai/openai-python: The official Python library for the OpenAI API}}.
\newblock \bibinfo{howpublished}{Available on \url{https://github.com/openai/openai-python}}.
\newblock


\bibitem[Plotkin and Power(2001)]%
        {FOSSACS:PP01}
\bibfield{author}{\bibinfo{person}{Gordon~D. Plotkin} {and} \bibinfo{person}{John Power}.} \bibinfo{year}{2001}\natexlab{}.
\newblock \showarticletitle{{Adequacy for Algebraic Effects}}. In \bibinfo{booktitle}{\emph{Foundations of Software Science and Computation Structures}} \emph{(\bibinfo{series}{FoSSaCS'01})}. \bibinfo{pages}{1--24}.
\newblock
\href{https://doi.org/10.1007/3-540-45315-6_1}{doi:\nolinkurl{10.1007/3-540-45315-6_1}}


\bibitem[Plotkin and Pretnar(2009)]%
        {ESOP:PP09}
\bibfield{author}{\bibinfo{person}{Gordon~D. Plotkin} {and} \bibinfo{person}{Matija Pretnar}.} \bibinfo{year}{2009}\natexlab{}.
\newblock \showarticletitle{{Handlers of Algebraic Effects}}. In \bibinfo{booktitle}{\emph{European Symp.\ on Programming}} \emph{(\bibinfo{series}{ESOP'09})}. \bibinfo{pages}{80--94}.
\newblock
\href{https://doi.org/10.1007/978-3-642-00590-9_7}{doi:\nolinkurl{10.1007/978-3-642-00590-9_7}}


\bibitem[Pretnar(2015)]%
        {ENTCS:Pretnar15}
\bibfield{author}{\bibinfo{person}{Matija Pretnar}.} \bibinfo{year}{2015}\natexlab{}.
\newblock \showarticletitle{{An Introduction to Algebraic Effects and Handlers (Invited tutorial paper)}}.
\newblock \bibinfo{journal}{\emph{Electr.\ Notes Theor.\ Comp.\ Sci.}}  \bibinfo{volume}{319} (\bibinfo{date}{December} \bibinfo{year}{2015}), \bibinfo{pages}{19--35}.
\newblock
\href{https://doi.org/10.1016/j.entcs.2015.12.003}{doi:\nolinkurl{10.1016/j.entcs.2015.12.003}}
\newblock
\shownote{The 31st Conference on the Mathematical Foundations of Programming Semantics (MFPS XXXI)}.


\bibitem[{Qwen Team}(2025)]%
        {arxiv:Qwen25}
\bibfield{author}{\bibinfo{person}{{Qwen Team}}.} \bibinfo{year}{2025}\natexlab{}.
\newblock \bibinfo{title}{{Qwen2.5 Technical Report}}.
\newblock
\href{https://doi.org/10.48550/arXiv.2412.15115}{doi:\nolinkurl{10.48550/arXiv.2412.15115}}


\bibitem[{\'S}cibior and Kammar(2015)]%
        {HOPE:SK15}
\bibfield{author}{\bibinfo{person}{Adam {\'S}cibior} {and} \bibinfo{person}{Ohad Kammar}.} \bibinfo{year}{2015}\natexlab{}.
\newblock \showarticletitle{{Effects in Bayesian inference}}. In \bibinfo{booktitle}{\emph{Workshop on Higher-Order Programming with Effects}} \emph{(\bibinfo{series}{HOPE'15})}.
\newblock


\bibitem[{The Guidance Contributors}(2025)]%
        {misc:Guidance25}
\bibfield{author}{\bibinfo{person}{{The Guidance Contributors}}.} \bibinfo{year}{2025}\natexlab{}.
\newblock \bibinfo{title}{{guidance-ai/guidance: A guidance language for controlling large language models.}}
\newblock \bibinfo{howpublished}{Available on \url{https://github.com/guidance-ai/guidance}}.
\newblock


\bibitem[Tran et~al\mbox{.}(2017)]%
        {ICLR:THS17}
\bibfield{author}{\bibinfo{person}{Dustin Tran}, \bibinfo{person}{Matthew~D. Hoffman}, \bibinfo{person}{Rif~A. Saurous}, \bibinfo{person}{Eugene Brevdo}, \bibinfo{person}{Kevin Murphy}, {and} \bibinfo{person}{David~M. Blei}.} \bibinfo{year}{2017}\natexlab{}.
\newblock \showarticletitle{{Deep Probabilistic Programming}}. In \bibinfo{booktitle}{\emph{Int.\ Conf.\ on Learning Representations}} \emph{(\bibinfo{series}{ICLR'17})}.
\newblock


\bibitem[Yao et~al\mbox{.}(2023)]%
        {NIPS:YYZ23}
\bibfield{author}{\bibinfo{person}{Shunyu Yao}, \bibinfo{person}{Dian Yu}, \bibinfo{person}{Jeffrey Zhao}, \bibinfo{person}{Izhak Shafran}, \bibinfo{person}{Thomas~L. Griffiths}, \bibinfo{person}{Yuan Cao}, {and} \bibinfo{person}{Karthik Narasimhan}.} \bibinfo{year}{2023}\natexlab{}.
\newblock \showarticletitle{{Tree of Thoughts: Deliberate Problem Solving with Large Language Models}}. In \bibinfo{booktitle}{\emph{Neural Info.\ Processing Syst.}} \emph{(\bibinfo{series}{NeurIPS'23})}. \bibinfo{pages}{11809--11822}.
\newblock
\urldef\tempurl%
\url{https://dl.acm.org/doi/abs/10.5555/3666122.3666639}
\showURL{%
\tempurl}


\bibitem[Yao et~al\mbox{.}(2025)]%
        {misc:ToT25}
\bibfield{author}{\bibinfo{person}{Shunyu Yao}, \bibinfo{person}{Dian Yu}, \bibinfo{person}{Jeffrey Zhao}, \bibinfo{person}{Izhak Shafran}, \bibinfo{person}{Thomas~L. Griffiths}, \bibinfo{person}{Yuan Cao}, {and} \bibinfo{person}{Karthik Narasimhan}.} \bibinfo{year}{2025}\natexlab{}.
\newblock \bibinfo{title}{{princeton-nlp/tree-of-thoughts-llm: [NeurIPS 2023] Tree of Thoughts: Deliberate Problem Solving with Large Language Models}}.
\newblock \bibinfo{howpublished}{Available on \url{https://github.com/princeton-nlp/tree-of-thought-llm}}.
\newblock


\bibitem[Zhao et~al\mbox{.}(2024)]%
        {ICML:ZBM24}
\bibfield{author}{\bibinfo{person}{Stephen Zhao}, \bibinfo{person}{Rob Brekelmans}, \bibinfo{person}{Alireza Makhzani}, {and} \bibinfo{person}{Roger Grosse}.} \bibinfo{year}{2024}\natexlab{}.
\newblock \showarticletitle{{Probabilistic Inference in Language Models via Twisted Sequential Monte Carlo}}. In \bibinfo{booktitle}{\emph{Int.\ Conf.\ on Machine Learning}} \emph{(\bibinfo{series}{ICML'24})}. \bibinfo{pages}{60704--60748}.
\newblock
\urldef\tempurl%
\url{https://dl.acm.org/doi/10.5555/3692070.3694582}
\showURL{%
\tempurl}


\bibitem[Zheng et~al\mbox{.}(2024)]%
        {arxiv:ZYX24}
\bibfield{author}{\bibinfo{person}{Lianmin Zheng}, \bibinfo{person}{Liangsheng Yin}, \bibinfo{person}{Zhiqiang Xie}, \bibinfo{person}{Chuyue Sun}, \bibinfo{person}{Jeff Huang}, \bibinfo{person}{Cody~Hao Yu}, \bibinfo{person}{Shiyi Cao}, \bibinfo{person}{Christos Kozyrakis}, \bibinfo{person}{Ion Stoica}, \bibinfo{person}{Joseph~E. Gonzalez}, \bibinfo{person}{Clark Barrett}, {and} \bibinfo{person}{Ying Sheng}.} \bibinfo{year}{2024}\natexlab{}.
\newblock \bibinfo{title}{{SGLang: Efficient Execution of Structured Language Model Programs}}.
\newblock
\urldef\tempurl%
\url{https://arxiv.org/abs/2312.07104}
\showURL{%
\tempurl}


\end{thebibliography}

\appendix
%\newpage
\pagebreak

\section{Towards Multi-shot Handlers}
\label{sec:multi-shot}

As we mentioned in \cref{sec:approach}, in theory, effect handlers do not need to be one-shot: a handler could resume the computation from invoking an operation multiple times.
A canonical usage is to implement \emph{nondeterminism} (as an operation) and \emph{backtracking} (as a handler).
Below is a pseudocode written in Python syntax for this usage.
The main difference is that each handling clause (e.g., lines 11 and 17) is parameterized by a specific parameter (e.g., \verb|resume| and \verb|unused_resume|), which is a function (also called a \emph{continuation}) that represents the computation from invoking an operation:
\begin{pythoncode}
# WARNING: THIS PSEUDOCODE WOULD NOT WORK!

decide = Operation()
fail = Operation()

class BacktrackHandler(Handler): # handles decide
  def decide(self, resume):
    class FailHandler(Handler):  # handles fail
      def fail(self, unused_resume):
        resume(False)
    with FailHandler():
      resume(True)

with BacktrackHandler():
  b1 = decide()
  b2 = decide()
  fail() if b1 ^ b2 else print(b1, b2)
\end{pythoncode}
In line 24, the operation \verb|decide| is dispatched to the clause defined at line 11, with the continuation \verb|resume| representing the computation from line 25 to line 26.
Lines 20 and 21 then push an \verb|FailHandler| and resume the computation with \mintinline{python}|True|, so the variable \verb|b1| gets the value \mintinline{python}|True|.
Similarly, in line 25, the operation \verb|decide| is dispatched with the continuation \verb|resume| representing the computation in line 26.
Also, another \verb|FailHandler| is pushed onto the stack, and the variable \verb|b2| gets the value \mintinline{python}|True|.
Because \verb|b1 ^ b2| is \mintinline{python}|False|, the \verb|fail| operation is invoked and discharged by the second \verb|FailHandler|.
According to lines 17 and 18, the computation on line 26 is resumed with \verb|b2| set to \mintinline{python}|False|.
This time, \verb|b1 ^ b2| becomes \mintinline{python}|True| and the execution finishes.

Backtracking, or more general searching, is a common pattern in LLM scripting, e.g., Tree-of-Thoughts~\cite{NIPS:YYZ23}.
However, due to the language's limitations, implementing multi-shot handlers in Python is not possible without modifying its runtime environment.
We leave this for future exploration.

\section{Semantics of Operations and Handlers}
\label{sec:semantics}

We did not discuss the implementation of \verb|Operation| (for abstract operations) and \verb|Handler| (for effect handlers) in \cref{sec:approach}.
Here, we give an operational semantics of a core calculus to demonstrate their intended behavior.
% and include the concrete implementation in the Appendix.

We consider the following core calculus, where $v$, $x$, $h$, $\m{op}$, $c$, and $\calH$ range over values, variables, handlers, operations, computations, and handler stacks, respectively.
Each handler $h$ specifies the set of operations it discharges and how to handle them.
For simplicity, we assume every operation receives exactly one argument.
In $\mwith{h}{c}$, we push $h$ onto the handler stack, proceed to execute $c$, and pop out $h$ after $c$ finishes.
We use $\varnothing$ to denote an empty stack and $h : \calH$ to denote a stack formed by pushing $h$ onto $\calH$.
\begin{align*}
  v & \Coloneqq x \mid \mathbf{true} \mid \mathbf{false} \mid \cdots \\
  h & \Coloneqq \handler{\m{op}_1(x) \mapsto c_1; \cdots; \m{op}_n(x) \mapsto c_n} \\
  c & \Coloneqq \mret{v} \mid \mdo{x}{c_1}{c_2} \mid \mif{v}{c_1}{c_2} \\
  & \,\,\mid\,\,\, \m{op}(v) \mid \mwith{h}{c} \\
  \calH & \Coloneqq \varnothing \mid h : \calH
\end{align*}

\begin{figure}[t]
\begin{mathpar}\small
  \Rule{LetStep}
  { \tuple{\calH; c_1} \leadsto \tuple{\calH; c_1'} }
  { \tuple{\calH; \mdo{x}{c_1}{c_2}} \leadsto \tuple{\calH; \mdo{x}{c_1'}{c_2}} }
  \and
  \Rule{LetBind}
  { }
  { \tuple{\calH; \mdo{x}{\mret{v_1}}{c_2}} \leadsto \tuple{\calH; [v_1/x]c_2} }
  \and
  \Rule{IfTrue}
  { }
  { \tuple{\calH; \mif{\mathbf{true}}{c_1}{c_2}} \leadsto \tuple{\calH; c_1} }
  \and
  \Rule{IfFalse}
  { }
  { \tuple{\calH; \mif{\mathbf{false}}{c_1}{c_2}} \leadsto \tuple{\calH; c_2} }
  \\
  \\
  \inferrule*[lab={(OpHandle)}]
  { h = \handler{\m{op}_1(x) \mapsto c_1; \cdots; \m{op}_n(x) \mapsto c_n} \\ 1 \le i \le n }
  { \tuple{h : \calH; \mdo{y}{\m{op}_i(v)}{c}} \leadsto \tuple{\calH; \mdo{y}{[v/x]c_i}{\mwith{h}{c}} }  }
  \and 
  \inferrule*[lab={(OpForward)}]
  { h = \handler{\m{op}_1(x) \mapsto c_1; \cdots; \m{op}_n(x) \mapsto x_n} \\ \m{op} \not\in \{ \m{op}_1,\cdots,\m{op}_n \} }
  { \tuple{h : \calH; \mdo{y}{\m{op}(v)}{c}} \leadsto \tuple{\calH, \mdo{y}{\m{op}(v)}{\mwith{h}{c}}} }
  \and
  \inferrule*[lab={(With)}]
  { }
  { \tuple{\calH; \mwith{h}{c}} \leadsto \tuple{h : \calH; c} }
\end{mathpar}
\caption{Rules of the operational semantics.}
\label{fig:semantics}
\end{figure}

\cref{fig:semantics} lists the rules of a small-step operational semantics for the core calculus.
The judgement $\tuple{\calH;c} \leadsto \tuple{\calH';c'}$ indicates a configuration with a handler stack $\calH$ and a computation $c$ can step to a new configuration $\tuple{\calH';c'}$.
Rule \textsc{(With)} takes a step from $\mwith{h}{c}$ by pushing $h$ to the handler stack.
Rule \textsc{(OpHandle)} deals with the case where the computation is $\mdo{y}{\m{op}_i(v)}{c}$ and the operation $\m{op}_i$ is discharged by the topmost handler.
We step from $\m{op}_i(v)$ to $[v/x]c_i$---a computation $c_i$ with $x$ substituted with $v$---because the handler $h$ provides the clause $\m{op}_i(x) \mapsto c_i$.
Note that we pop $h$ from the stack during processing $c_i$ but push it back when executing $c$.
Rule \textsc{(OpForward)} deals with the case where the computation is $\mdo{y}{\m{op}(v)}{c}$ but the topmost handler does not discharge the operation $\m{op}$.
We keep $\m{op}(v)$ unchanged but pop $h$ from the stack, attempting to discharge $\m{op}$ in the remaining stack.
Similarly to \textsc{(OpHandle)}, we need to push $h$ back when executing $c$.

\end{document}